\documentclass[aps,prb,twocolumn,notitlepage,showpacs,superscriptaddress,am]{revtex4-1}%
\usepackage{graphicx}
\usepackage{amsmath}
\usepackage{amssymb}
\usepackage{color}
\usepackage{amsfonts}%
\begin{document}
\title{Transition of a prestine Mott insulator to a correlated Fermi liquid:\\
Pressure-dependent optical investigations of a quantum spin liquid}

\author{Weiwu Li}
\affiliation{1.~Physikalisches Institut, Universit\"at Stuttgart, Pfaffenwaldring 57,
70569 Stuttgart, Germany}
\author{Andrej Pustogow}
\affiliation{1.~Physikalisches Institut, Universit\"at Stuttgart, Pfaffenwaldring 57,
70569 Stuttgart, Germany}
\author{Reizo Kato }
\affiliation{RIKEN, 2-1, Hirosawa, Wako-shi, Saitama 351-0198, Japan}
\author{Martin Dressel}
\affiliation{1.~Physikalisches Institut, Universit\"at Stuttgart, Pfaffenwaldring 57,
70569 Stuttgart, Germany}
\date{\today}

\begin{abstract}
Systematic pressure- and temperature-dependent infrared studies on the two-dimensional organic quantum spin-liquid $\beta^{\prime}$-EtMe$_3$Sb[Pd(dmit)$_2$]$_2$
disclose the electronic and lattice evolution across the Mott insulator-metal transition.
Increasing hydrostatic pressure continuously suppresses the insulating ground state; for $p>0.6$~GPa, a Drude-like component develops
indicating the appearance of coherent quasiparticles at the Fermi level.
In the vicinity of the Mott transition, not only the electronic state changes rapidly, but also the vibration modes exhibit a jump both in frequency and Fano constant, underlining the strong coupling between lattice and electrons.
The anisotropy of the in-plane optical response becomes inverted above 0.6~GPa. The findings are discussed in detail and summarized in a phase diagram comprising different experimental approaches.
\end{abstract}

\maketitle

\section{Introduction}
The physics governing the Mott metal-insulator transition
is of paramount importance for understanding strongly correlated electron systems.
It is based on the concept that the electrons in
a half-filled system -- i.e. one charge per lattice site -- tend to localize
when electron-electron interactions become strong.
From the viewpoint of the simple Hubbard model, the key parameter to control the metal-insulator transition is the ratio of Coulomb repulsion $U$ to bandwidth $W$.
Despite enormous progress in dynamical mean-field theory (DMFT) \citep{Georges96,Kotliar06} and
decades of investigating typical Mott insulators, such as
V$_{2}$O$_{3}$ \cite{Mcwhan73,Imada98,Limelette03,Arcangeletti07},
their ground state and
the correlation-driven phase transition remain a challenge to condensed matter physicists because
antiferromagnetic order  -- commonly observed in these compounds at low temperatures \cite{Imada98} -- obscures the genuine Mott state.

The solution is offered by quantum spin liquids, i.e.\ Mott insulators that do not exhibit any sign of long-range magnetic order despite  strong antiferromagnetic coupling \cite{Norman16,Savary17,Zhou17,Dressel18}.
Several quasi-two-dimensional molecular crystals came under particular scrutiny as they form highly frustrated triangular arrangements of dimers occupied by a single charge with $S=\frac{1}{2}$.
In $\kappa$-(BEDT-TTF)$_2$\-Cu$_2$(CN)$_3$ or $\beta^{\prime}$-EtMe$_3$\-Sb\-[Pd(dmit)$_2$]$_2$, for instance, $J \approx 200$~K but no ordering occurs down to a few mK \cite{Kanoda11,Zhou17}.
When temperature and correlations are scaled by the bandwidth: $T/W$ and $U/W$, a generic phase diagram of the Mott insulator-metal transition
is unveiled, with the quantum Widom line indicating the crossover from bad metal to Mott insulator, a critical endpoint of the phase boundary followed by a Pomeranchuk-like back-bending and metallic quantum fluctuations in a coexistence regime \cite{Pustogow17}.
However, not much is known about how the dynamics of charge carriers develops as correlations advance across the metal-insulator transition.
What is the nature of the metallic state that gradually evolves in these highly frustrated compounds?

For the rather soft compounds, hydrostatic pressure is the most suitable way to tune across the insulator-metal phase boundary.
When pressurized by only 0.6 GPa, for instance, $\beta^{\prime}$-EtMe$_3$Sb\-[Pd(dmit)$_2$]$_2$ becomes metallic at low temperatures without signs of superconductivity \cite{Itou17}.
Optical spectroscopy is the superior method for investigating the electronic properties of organic conductors,
because it allows us to analyse the dynamics of the charge carriers and directly extract the Coulomb repulsion $U$ and the bandwidth $W$ as well as the coherent quasiparticle contribution to the dynamical conductivity.
Here we want to explore in detail the transition from the Mott insulating to the metallic state by performing pressure- and temperature-dependent optical investigations on $\beta^{\prime}$-EtMe$_3$Sb\-[Pd(dmit)$_2$]$_2$ single crystals.
We see how the Mott gap is suppressed on the insulating side and how
the effective mass evolves on the metallic side.
A thorough analysis of the vibrational features provides interesting insight into the electron-phonon coupling and changes upon crossing the insulator-metal transition.


\section{Experimental Details}
\begin{figure}
\centering
\includegraphics[width=1\columnwidth]{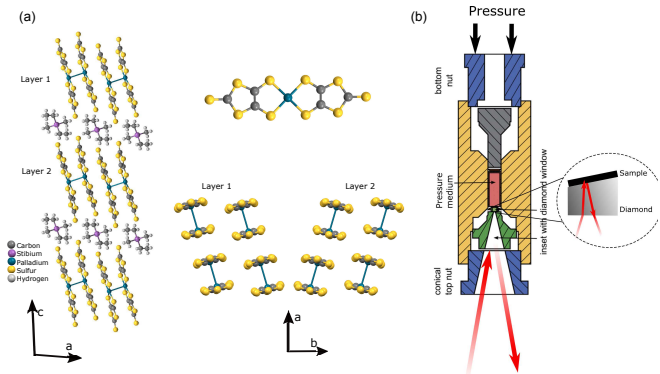}%
\caption{(a) Illustration of the $\beta^{\prime}$-EtMe$_3$Sb[Pd(dmit)$_2$]$_2$ molecular and crystal structure, consisting of two crystallographically equivalent Pd(dmit)$_2$ layers (layer 1 and layer 2) separated by the counterions EtMe$_3$Sb.
(b)~Sketch of the piston pressure cell utilized for pressure-dependent infrared studies. The reflectivity is measured off the sample-diamond interface; the outer diamond surface is used as reference. For more details on the sample preparation, optical technique and analysis see \cite{Beyer15}.}
\label{crystal_cell}
\end{figure}

High quality single crystals of $\beta^{\prime}$-EtMe$_3$Sb[Pd(dmit)$_2$]$_2$ are prepared by an aerial oxidation method as described previously \cite{Kato12}; here
EtMe$_3$Sb stands for ethyl-trimethyl-stibonium and dmit is 1,3-dithiole-2-thione-4,5-dithiolate.
As shown in Fig.~\ref{crystal_cell}(a) the organic conductor $\beta^{\prime}$-EtMe$_3$Sb[Pd(dmit)$_2$]$_2$ consists of alternating conducting layers of [Pd(dmit)$_2$]$_2^-$ dimers and insulating layers of EtMe$_3$Sb$^+$ cations.
The quasi-two-dimensional electronic properties stem from the conduction band formed by the dominant $p$-orbital of the sulphur atom in [Pd(dmit)$_2$]$_2$.
The bands of EtMe$_3$Sb are well below the Fermi level \cite{Scriven12,Nakamura12,Ueda18}.

Specimens with shiny and flat surfaces are selected for optical measurements; the large size up to 0.5~mm allows measuring the low-energy response down to far-infrared frequencies.
High-pressure experiments from $p=0.23$ to 1.2~GPa were conducted in a clamp-type BeCu cell [Fig.~\ref{crystal_cell}(b)] with Daphne 7373-type oil as pressure medium \cite{Beyer15}.
With the help of a home-built He-cryostat attached to a Fourier-transform spectrometer, temperature-dependent reflectivity measurements
spanning from 150 to 8000~cm$^{-1}$ were performed down to $T=10$~K. The use of proper polarizers allow us to probe the response along different crystal directions. The optical conductivity is calculated via a Kramers-Kronig analysis with a Hagen-Rubens extrapolation for the metallic state and a constant low-frequency reflectivity in the insulating state \cite{DresselGruner02}.


\section{Results and Analysis}
\begin{figure*}
\centering
\includegraphics[width=0.8\textwidth]{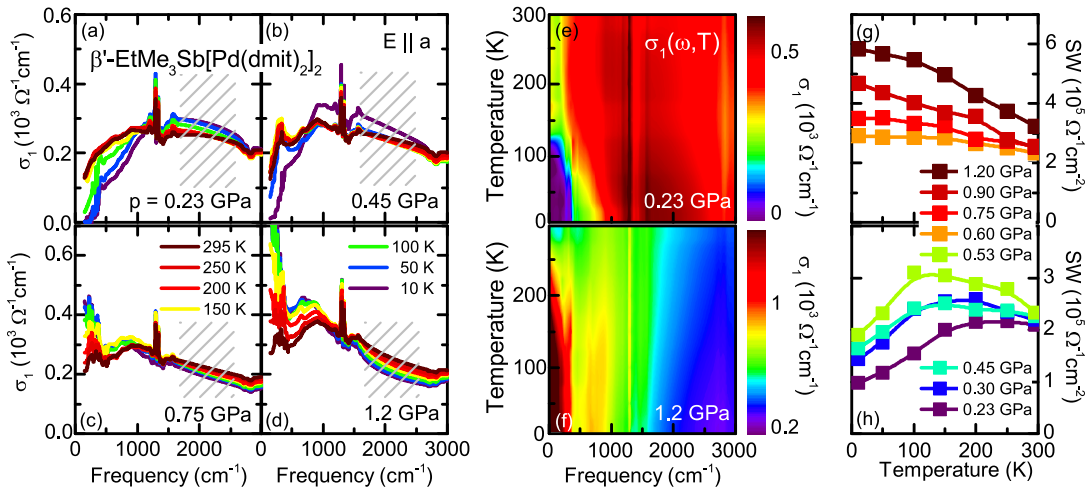}%
\caption{(a)-(d)~Temperature evolution of the $a$-axis optical conductivity of $\beta^{\prime}$-EtMe$_3$Sb[Pd(dmit)$_2$]$_2$ for selected hydrostatic pressures below and above the Mott transition at $p_c=0.6$~GPa.
The shaded area indicates the strong light absorption by the diamond window preventing reliable measurements; we linearly interpolate the optical reflectivity in this frequency range.
(e)~False color plot of the temperature-dependent conductivity spectra $\sigma_1(\omega,T)$ for $p=0.23$~GPa ranging to $\sigma_1 = 600~\Omega^{-1}{\rm cm}^{-1}$. (f)~Contour plot $\sigma_1(\omega,T)$ for $p=1.2$~GPa spanning from $\sigma_1 = 200~\Omega^{-1}{\rm cm}^{-1}$ (dark blue) to $1500~\Omega^{-1}{\rm cm}^{-1}$ (dark red). In the insulating state ($p=0.23$~GPa), the spectrum is dominated by the mid-infrared band with a clear gap opening upon cooling; while for $p=1.2$~GPa, a metallic behavior with a Drude-like contribution becomes obvious.
(g)-(h) The temperature dependence of the low-frequency spectral weights integrated up to 700~cm$^{-1}$ exhibit different behaviors for $p<p_c$ compared to (g)~high pressure, $p>p_c$.}
\label{Fig:Cond_T}%
\end{figure*}

The temperature dependence of the optical properties measured
with light polarized parallel to the $a$-axis is presented in
Fig.~\ref{Fig:Cond_T} for selected pressures below and above the critical pressure $p_c$, defining the insulator-metal transition.
A qualitatively similar temperature behavior is observed for $E\parallel b$; the complete set of spectra is presented in the Supplemental Materials \cite{SM}.
Under ambient conditions the optical conductivity
contains a pronounced mid-infrared absorption around 2000~cm$^{-1}$
and a non-vanishing zero-frequency conductivity due to some incoherent bad metallic conductivity; a behavior rather typical for most families of organic metals \cite{Dressel04,Pustogow17}.
The two strong and sharp vibrational features observed around
1200~cm$^{-1}$ result from electron-molecular vibrational modes and have been extensively investigated by Yamamoto {\it et al.} \cite{Yamamoto11a,Yamamoto11b,Yamamoto17}.
As the temperature is reduced, the low-frequency conductivity first rises slightly, but below 200~K it gradually vanishes as the Mott gap develops.  The overall spectrum and temperature behavior does not change substantially when pressure increases up to $p=0.6$~GPa.

From the false-color contour plots of the $p=0.23$ and 1.2~GPa data, we see the qualitatively different behavior in the insulating and metallic regimes.
For a quantitative analysis of the Drude-like contribution, the temperature dependence of the spectral weight $SW= \int^{700}_{0}\sigma_1(\omega){\rm d}\omega$ is plotted in Fig.~\ref{Fig:Cond_T}(g) and (h).
In the insulating state $SW$ first rises upon lowering the temperature, but then it decreases as the Mott gap in the charge excitations gradually opens. This transition temperature is reduced with pressure,
indicating a transition from a insulator-like to a metal-like behavior.
Our conclusion is corroborated by pressure-dependent transport measurements \cite{Itou17}, where the change in slope from ${\rm d}\rho/{\rm d}T>0$ to ${\rm d}\rho/{\rm d}T<0$ determines the crossover from metal to insulator.
Above $p_c = 0.6$~GPa the spectral weight exhibits a completely different temperature behavior: the continuous increases down to $T=10$~K is a measure for  the build-up of the coherent charge response.
We do not see indication of a complete energy gap or pseudogap in our data, as inferred from the low-temperature upturn in the high-pressure resistivity \cite{Itou17};
future experiments at lower temperatures and smaller frequencies might clarify this controversy.
It is interesting to note the recent results on a low-energy gap in the Dirac system $\alpha$-(BEDT-TTF)$_{2}$I$_{3}$ where strong correlations eventually become effective \cite{Liu16,Uykur18}.


\begin{figure}[h]
\centering
\includegraphics[width=0.9\columnwidth]{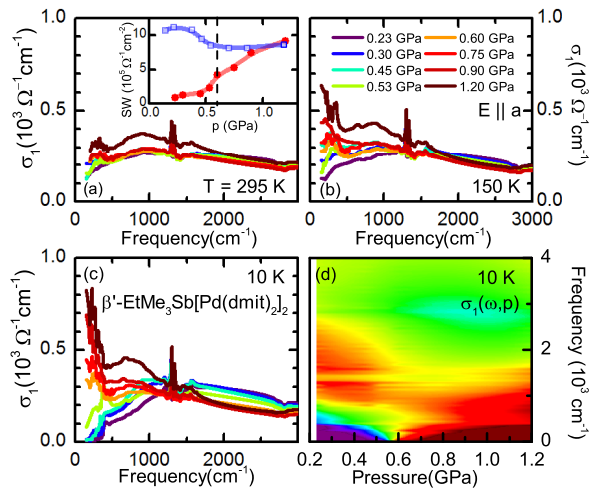}%
\caption{(a)-(c)~Pressure dependence of the optical conductivity of $\beta^{\prime}$-EtMe$_3$Sb[Pd(dmit)$_2$]$_2$  for selected temperatures $T=295$, 150 and 10~K.
The inset displays the pressure evolution of the spectral weight for the low-energy (red dots) and mid-infrared region (blue squares) at $T=10$~K; the dashed line indicate the critical pressure $p_c$ of the Mott transition.
(d)~In the contour plot $\sigma_1(\omega,p)$, the variations of the low-temperature conductivity spectra with pressure are presented. The color code is similar to Fig.~\ref{Fig:Cond_T}(f). The Mott insulator-metal transition at $p_c=0.6$~GPa is clearly visible.}%
\label{Fig:Cond_p}%
\end{figure}

In order to demonstrate the pressure dependence of the optical conductivity
more clearly, in Fig.~\ref{Fig:Cond_p} we present the spectra recorded at $T=295$, 150 and 10~K.
At room temperature the mid-infrared absorption is barely affected by pressure, while the low-energy range increases slightly without developing a pronounced Drude peak; $\beta^{\prime}$-EtMe$_3$Sb[Pd(dmit)$_2$]$_2$ remains in a bad metallic state over the whole pressure range. Our findings are in accord with dc measurements \cite{Itou17} indicating a decrease of resistivity within factor of ten as pressure increases to 1.8~GPa.
The difference between low and high-pressure spectra becomes more pronounced
when the sample is cooled down to $T=150$~K, as shown in Fig.~\ref{Fig:Cond_p}(b).
In particular below 500~cm$^{-1}$ the behavior clearly forks:
a sharp Drude peak develops for $p=0.9$ and 1.2~GPa, while the far-infrared conductivity is suppressed in the case of $p=0.23$~GPa due to the gradual opening of the Mott gap upon cooling; this is reflected in the temperature behavior of the spectral weight, plotted in Fig.~\ref{Fig:Cond_T}(h).

When going to the lowest temperature, thermally excited charge carriers freeze out completely; at  $T = 10$~K a full gap has developed in the Mott state ($p_c < 0.6$~GPa), while above $p_c$ a Drude peak is present that becomes enhanced with increasing pressure. The corresponding spectral-weight transfer is quantitatively analyzed in the inset of Fig.~\ref{Fig:Cond_p}(a), where the Drude term and mid-infrared band $\int^{\omega_u}_{\omega_l}\sigma_1(\omega){\rm d}\omega$,
(with the lower and upper bound $\omega_l= 1000~{\rm cm}^{-1}$ and $\omega_u= 3000~{\rm cm}^{-1}$, respectively)
is plotted as a function of pressure.
The small $SW$ at low-energies increases significantly only, when
pressure exceeds $p_c$ and the Mott gap has closed.
Concomitantly the mid-infrared $SW$ drops rapidly when the phase boundary is crossed.
The transfer of $SW$ over such a large energy scale is taken as evidence of strong correlations \cite{DresselGruner02,Basov11}.
Fig.~\ref{Fig:Cond_p}(d) illustrates that the change of optical conductivity at $T=10$~K is most severe when pressure increases from 0.53 to 0.6~GPa. Here the Mott gap closes and the compound becomes metallic for higher pressure.


\section{Discussion}
\subsection{Electron-Electron Interaction}

\begin{figure}
\centering
\includegraphics[width=1.0\columnwidth]{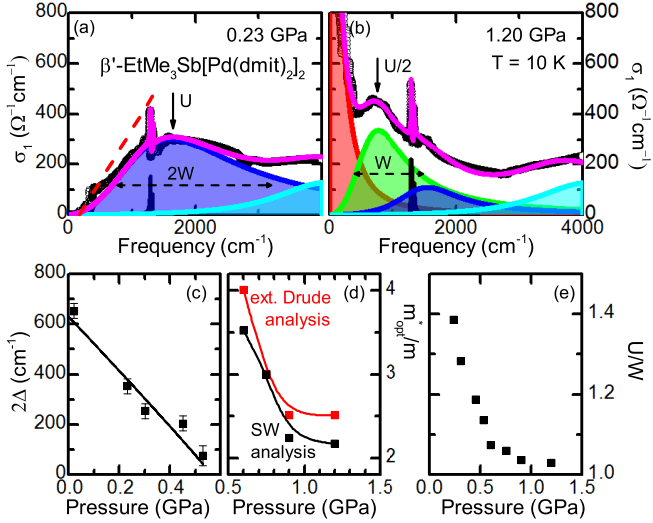}%
\caption{(a) and (b)~Drude-Lorentz fit to the 0.23 and 1.2~GPa spectra of $\beta^{\prime}$-EtMe$_3$Sb[Pd(dmit)$_2$]$_2$ measured at $T=10$~K.
The excitation across the Hubbard band are described by a band (blue)  centered at $U$ and with half-width $W$, higher energy contributions are given an additional oscillator (cyan). The metallic phase is characterized by a pronounced Drude contribution (red)
and excitations between these itinerant carriers to the Hubband band and {\it vice versa} (green oscillator at $U/2$). In the insulating case the linear extrapolation (dashed line) is used to determine the Mott gap. An additional Fano term (black) is used to describe the vibrational feature. The sum of these contributions (magenta) fit the data (dots) well.
(c)~The extracted Mott gap decreases linearly with pressure.
(d)~The effective mass $m^{*}/m_b$ becomes strongly enhanced when $p_c$ is approached from the metallic side.
The red squares are calculated by the extended Drude analysis,
given in Eq.~(\ref{eq:m_eD});
the black squares are obtained from the spectral weight analysis, Eq.~(\ref{eq:m_SW}).
(e)~Pressure evolution of the effective correlations $U/W$ as extracted from the low-temperature spectra.}
\label{Fig:Analysis}%
\end{figure}
In a next step we performed a Drude-Lorentz analysis of our spectra, in order to directly compare the results with theoretical predictions \cite{Georges96,Kotliar06,Merino08}.
As examples, the low-temperature conductivity for the lowest and highest pressure is plotted in Fig.~\ref{Fig:Analysis} together with the corresponding fit \cite{SM}.
For the insulating state ($p=0.23$~GPa) a satisfactory description is reached with one broad mid-infrared band (dark blue) centered at 1500~cm$^{-1}$
and an additional high-energy tail (cyan) accounting for the interband transitions.
At $p=1.2$~GPa two additional contributions are necessary: a Drude
term and an additional mode at 750~cm$^{-1}$.
According to single-band DMFT calculated in the realm of the single-band Hubbard model \cite{Rozenberg95,Georges96}
these features are assigned to transitions between the lower and upper Hubbard bands separated by $U$; the width of the spectral feature corresponds to twice the bandwidth $2W$.
The contributions of the coherent quasi-particles at the Fermi energy, and the excitations between the Hubbard band and the central peak at $U/2$ appear only when $U/W$ drops below a critical value $(U/W)_{c}$.
In Fig.~\ref{Fig:Analysis}(e) the effective correlations $U/W$ are plotted as a function of pressure. While the onsite Coulomb repulsion $U$ varies only slightly with pressure, the bandwidth increases significantly  \cite{SM}. From our experiments we estimate $(U/W)_{c}\approx 1.1$ in excellent agreement with DMFT predictions \cite{Vucicevic13}.

In the insulating phase, we can estimate the charge gap via linear extrapolation as depicted by the dashed line in Fig.~\ref{Fig:Analysis}(a).
In a first approximation it
linearly decreases with pressure from the ambient pressure value 2$\Delta_0=600~{\rm cm}^{-1}$ until it closes at $p_c=0.6$~GPa.

The degree of electronic correlation in the metallic state is commonly expressed by the effective mass $m^{*}$, which can be estimated from the analysis of the spectral weight
$SW = \int\sigma_1(\omega){\rm d}\omega = \omega_p^2/8 = \frac{\pi}{2}\frac{Ne^2}{m^*}$.
Already the inset of Fig.~\ref{Fig:Cond_p}(a) illustrates that
the overall spectral weight is not recovered within the infrared range considered; this loss of $SW$ with decreasing pressure
evidences the enhancement of correlations.
For a quantitative analysis we consider the ratio of the zero-frequency term $\int\sigma_{\rm Drude}(\omega){\rm d}\omega$ related to  the itinerant carriers and the intraband contribution obtained by subtracting the interband contributions from the total conductivity,
$SW_{\rm intra} = SW_{\rm total} - SW_{\rm inter}$:
\begin{equation}
\frac{m^{*}_{\rm SW}}{m_b} = \frac{\int\sigma_1^{\rm intra}(\omega){\rm d}\omega} {\int\sigma_1^{\rm Drude}(\omega){\rm d}\omega } \quad .
\label{eq:m_SW}
\end{equation}
The development of ${m^{*}_{\rm SW}}/{m_b}$ with pressure is plotted in Fig.~\ref{Fig:Analysis}(d) as black squares.

Another approach starts from Fermi-liquid theory and considers
the energy-dependent scattering rate and effective mass.
From the extended Drude analysis of the complex optical conductivity $\hat{\sigma}(\omega)=\sigma_{1}(\omega)+{\rm i} \sigma_{2}(\omega)$  we obtain \cite{DresselGruner02}
\begin{equation}
\frac{m^{*}(\omega)}{m_b} = \frac{\omega^{2}_{p}}{4\pi\omega} \frac{\sigma_{2}(\omega)}{\sigma_{1}^{2}(\omega)+\sigma_{2}^{2}(\omega)} \quad ,
\label{eq:m_eD}
\end{equation}
where the plasma frequency $\omega^{2}_{p}=8\int^{\omega_{c}}_{0}\sigma_1(\omega){\rm d}\omega$ is calculated up to a cutoff  $\omega_{c}=3000~{\rm cm}^{-1}$, chosen to disregard interband transitions.
\begin{figure}
\centering
\includegraphics[width=0.6\columnwidth]{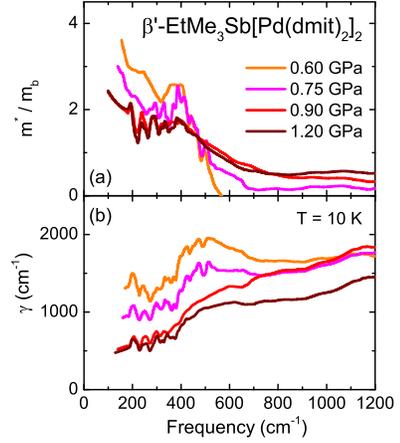}%
\caption{Frequency dependence of the (a) effective mass and (b) scattering rate of $\beta^{\prime}$-EtMe$_3$Sb[Pd(dmit)$_2$]$_2$ as determined by the extended Drude analysis of the low-temperature optical reflectivity measured at different pressure values in the metallic state. }
\label{Fig:extDrude}%
\end{figure}
For $T=10$~K, we plot the frequency dependence of the effective mass $m^*(\omega)/m_b$ and scattering rate $\gamma(\omega)$ for different pressures, $p>p_c$, in Fig.~\ref{Fig:extDrude}. While at high energies the effective mass is basically energy independent, we see a prominent increase for frequencies below $600~{\rm cm}^{-1}$. This dispersion becomes more pronounced as the Mott insulator transition is approached with lowering pressure. Correspondingly the scattering rate $\gamma(\omega)$ decreases as we go down in frequency, as displayed in Fig.~\ref{Fig:extDrude}(b). The quality of the data and systematic errors do not allow us to draw quantitative conclusions on the functional dependence.

In Fig.~\ref{Fig:Analysis}(d) the effective mass is plotted as a function of pressure obtained from the spectral weight evaluation [Eq.~(\ref{eq:m_SW})] and the $\omega\rightarrow 0$ limit of the extended Drude analysis [Eq.~(\ref{eq:m_eD})]. Both results are in good accord as far as the pressure dependence and the absolute value is concerned. When $p$ exceeds the critical value $p_c$, the system becomes less correlated and $m^{*}/m_b$ decreases.
The corresponding analysis of the $E\parallel b$ spectra yield a similar behavior.
Our findings are also in accord with observations reported from optical investigations
on $\kappa$-phase BEDT-TTF salts tuned by physical or chemical pressure \cite{Klehe00,McDonald03,Merino08,Dumm09}. As the systems approach the metal-to-insulator transition from the metallic side by enhancing electronic correlations, the effective mass increases rapidly.

\subsection{Electron-Phonon Interaction}

\begin{figure}
\centering
\includegraphics[width=0.9\columnwidth]{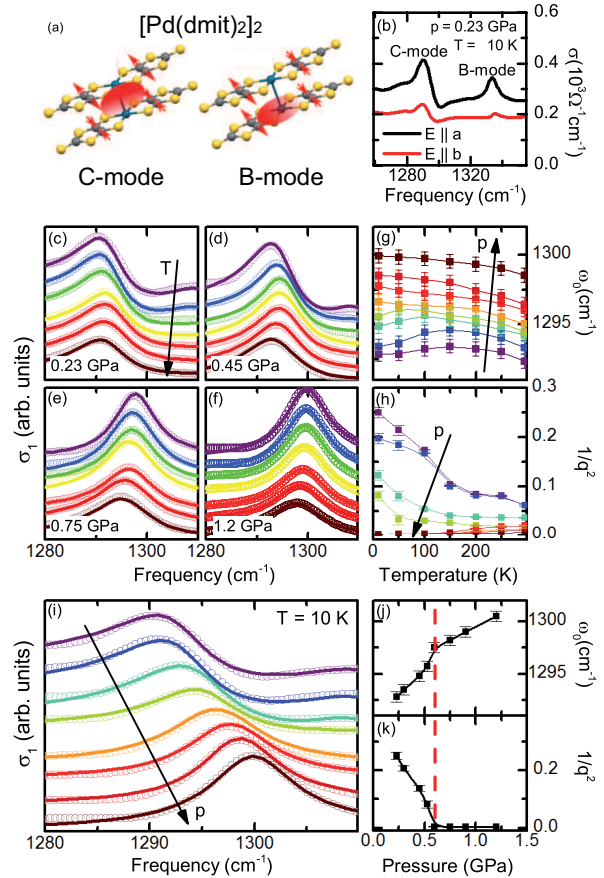}%
\caption{(a)~Schematic view of infrared-active C=C stretching modes of [Pd(dmit)$_2$]$_2$ dimers. The motion of the carbon atoms is indicated by red arrows. The gradual red color depicts the local electron density, which induces a dipole moment. The electric dipole moments of the C-mode and B-mode are perpendicular to the dimer (inter-molecular) and along the Pd(dmit)$_2$ molecular axis (intra-molecular), respectively. (b)~ C- and B-modes in the optical conductivity spectra of both polarizations recorded at $p = 0.23$~GPa and  $T=10$~K. (c)-(f)~Temperature evolution of the C-mode features and corresponding Fano fits. The data obtained from $T=295$ down to 10~K are presented for selected pressures, as indicated. Panels (g) and (h) display the temperature dependence of the fit parameters: (g) the central frequency $\omega_{0}$ and (h) the Fano
coupling constant $1/q^{2}$ for all measured pressures from $p=0.23$ to 1.20~GPa. (h)~The pressure dependence of $T=10$~K spectra shows the hardening of the mode as pressure increases and the remarkable change in shape; the solid lines correspond to Fano fits with parameters displayed in panels (j) and (k).
Both the center frequency $\omega_0$ and coupling constant $1/q^2$ exhibit an anomaly at the Mott transition indicated by the red dashed line.}
\label{Fig_Vibrations}%
\end{figure}

Ideally the Mott transition is supposed to represent a purely electronic phase transition; in a real crystals, however, a coupling to the underlying lattice is unavoidable. Optical spectroscopy provides the unique tool to investigate the vibrational response of the molecular and crystal structure at the phase transition in order to extract information on changes in structure and charge distribution.
Fig.~\ref{Fig_Vibrations} displays the behavior of the two strongest infrared-active A$_g$ vibrational modes of $\beta^{\prime}$-EtMe$_3$Sb[Pd(dmit)$_2$]$_2$ as a function of temperature and pressure.
The four C=C bonds in the Pd(dmit)$_2$ molecular dimer vibrate such that they are in phase within one molecule but out of phase with respect to the sibling molecule (C-mode); the B-mode denotes the out-of-phase vibration within the molecules as depicted in panel (a) \cite{Yamamoto17}.
Since the electric dipole is perpendicular to the molecular axis, the C-mode couples stronger to the electronic background compared to the B-mode where the dipoles point along the Pd(dmit)$_2$ axes \cite{Yamamoto11a,Yamamoto11b,Yamamoto17}.
Due to thermal contraction, the vibrational features in general harden upon cooling; for lowest pressure, however, we observe a blue shift at the Mott transition and the Fano shape becomes more pronounced [Fig.~\ref{Fig_Vibrations}(c)-(h)]. We interpret this behavior as indication that the electronic screening changes upon approaching the metal-insulator transition.

The low-temperature spectra exhibit a clear change in frequency and electronic coupling when the pressure drops below $p_c$, as depicted in Fig.~\ref{Fig_Vibrations}(j)-(k).
It is interesting to note that thermal expansion measurements on the related compound $\kappa$-(BEDT-TTF)$_2$Cu[N(CN)$_2$]Cl also reveal a jump of the expansion coefficient when tuned by pressure across the first-order insulator-metal line \cite{Gati16}.

Our optical investigations now give conclusive evidence that the strength of itinerant (inter-dimer) charge contribution  increase with pressure but on the expense of the localized (intra-dimer) spectral weight. Above $p_c=0.6$~GPa the vibrational modes become decoupled from the electronic background.


\subsection{Electronic Anisotropy}

\begin{figure}%
\centering
\includegraphics[width=1\columnwidth]{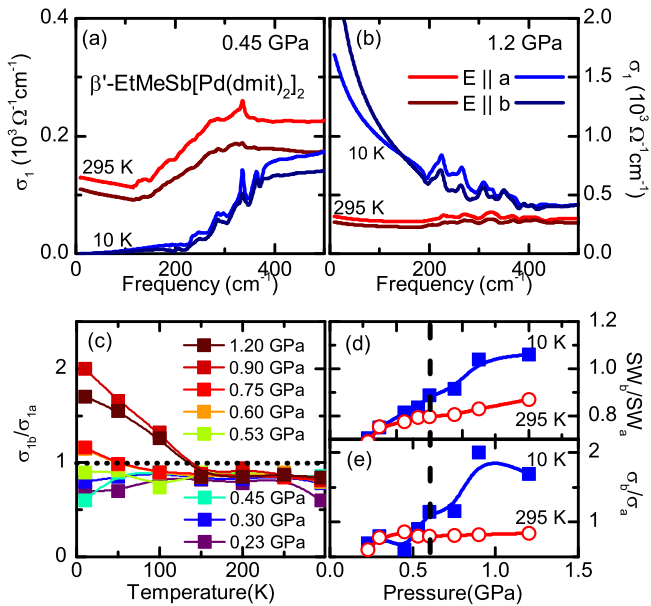}%
\caption{(a,b) Conductivity spectra of $\beta^{\prime}$-EtMe$_3$Sb[Pd(dmit)$_2$]$_2$
recorded for two different polarizations, $E\parallel a$ and $E\parallel b$, at room temperature and $T=10$~K.  (a) For low pressure $p<p_c$ the anisotropy is conserved with temperature, (b) while a reversal at low-frequencies is observed for $p>p_c$.
(c)~Temperature dependence of anisotropy ratio of the conductivity $\sigma_{1b}/\sigma_{1a}$ in the $\omega\longrightarrow$0 limit at various pressure values. In the metallic state the Hagen-Rubens extrapolated value are used, for the insulating state we took the value just above the Mott gap since inside the gap the conductivity drops to almost zero. The dotted line indicates the isotropy $\sigma_b$ = $\sigma_a$. (d) and (e) show the pressure dependent anisotropy of the spectral weight $SW$ and conductivity($\omega\rightarrow 0$) for $T=295$ and 10~K. The dashed line indicates the critical pressure $p_c=0.6$~GPa of Mott transition}
\label{Fig_anistropy}%
\end{figure}
$\beta^{\prime}$-EtMe$_3$Sb[Pd(dmit)$_2$]$_2$ is a quasi-two-dimensional conductor with a small anisotropy at ambient conditions: $\sigma_{1a} > \sigma_{1b}$.
While the ratio $\sigma_{1b}/\sigma_{1a} = 0.8$ to 0.9 remains constant throughout the insulating state ($p<p_c$),
the Drude term is found to increase more rapidly for $E\parallel b$ than for $E\parallel a$ when entering the metallic phase. In Fig.~\ref{Fig_anistropy} we present the temperature and pressure dependence of the anisotropy $\sigma_{1b}/\sigma_{1a}$ as obtained from the $\omega \rightarrow 0$ data indicating the inversion for low temperature and high pressure. Similar results are obtained from the analysis of the spectral weight of the Drude term in both directions. Our findings infer the opening of a pseudogap along the $a$-axis.
Beside investigations of the thermal expansion, we suggest more detailed studies of the Fermi surface of $\beta^{\prime}$-EtMe$_3$Sb[Pd(dmit)$_2$]$_2$ for high pressure and low temperature where the Mott phase transition is crossed and the system turns from an insulator to a metal.
Hall measurements indicate an electron-hole doping asymmetry in the antiferromagnetic Mott insulator $\kappa$-(BEDT-TTF)$_2$Cu[N(CN)$_2$]Cl \cite{Kawasugi16}.
Recent Hall studies  on the $\kappa$-salt spin-liquid Mott insulators $\kappa$-(BEDT-TTF)$_2X$ yield evidence for Mott-Anderson localization \cite{Culo19}. It would be of interest to extend these Hall investigations
to the title compound $\beta^{\prime}$-EtMe$_3$Sb[Pd(dmit)$_2$]$_2$.


\section{Summary and Conclusions}
Employing temperature- and pressure-dependent optical investigations on the quantum spin liquid candidate $\beta^{\prime}$-EtMe$_3$Sb[Pd(dmit)$_2$]$_2$, we could elucidate the electrodynamics at the Mott insulator-metal  transition in detail.
The results unambiguously confirm the realization of the pure bandwidth-controlled phase transition via applied pressure. From our optical data we can determine values for the onsite Coulomb repulsion $U$ and the bandwidth $W$.
With increasing pressure the effective correlations $U/W$ are continuously reduced.
The extracted parameters of $T/W$ and $U/W$ are in good agreement with DMFT calculations \cite{SM}.
\begin{figure}[b]
\centering
\includegraphics[width=1\columnwidth]{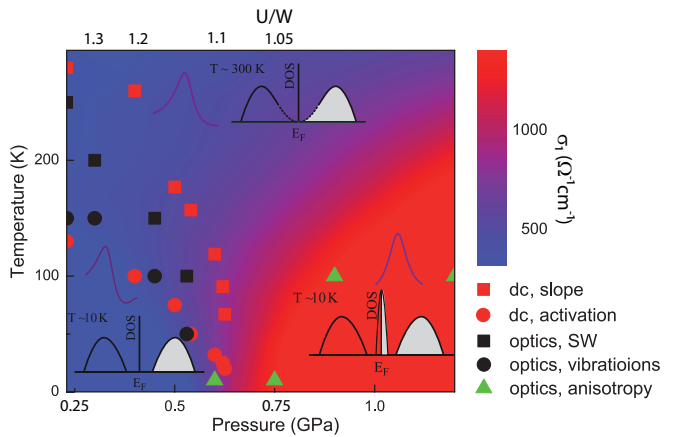}%
\caption{Schematic pressure-temperature  diagram of $\beta^{\prime}$-EtMe$_3$Sb[Pd(dmit)$_2$]$_2$. The false-coloured contour plot is based on the optical conductivity along the $a$-axis in the limit $\omega\rightarrow 0$.
The red squares refer to the change in the slope of the dc transport \cite{Itou17} where the
temperature of the metal-insulator transition is defined by ${\rm d}\rho/{\rm d}T=0$.
The red dots indicate the temperatures, at which the activation energy
changes; i.e.\ the maximum of ${\rm d\,ln}\rho/{\rm d}(1/T)$.
From our optical experiments we analyse the temperature dependence of the vibration features
and shift in spectral weight.
The black dots corresponds to the maxima of the C-mode frequency plotted in Fig.~\ref{Fig_Vibrations}(g). The black squares refer to the temperatures where the low-frequency spectral weight is largest, as plotted in Fig.~\ref{Fig:Cond_T}. The green triangles represent the temperatures, when the ratio of $\sigma_{1b}$/$\sigma_{1a}$ crosses unity. The upper scale is extracted from Fig.~\ref{Fig:Analysis}(e). See Supplemental Materials \cite{SM} for more details.}
\label{phase_diag}%
\end{figure}

In the pressure-temperature phase diagram of Fig.~\ref{phase_diag} we summarize our optical experiments and complement them with transport data \cite{Itou17}.
At ambient pressure $\beta^{\prime}$-EtMe$_3$Sb[Pd(dmit)$_2$]$_2$ is a Mott insulator with a
Mott-Hubbard gap of $600~{\rm cm}^{-1}$ that linearly decreases with pressure and vanishes at $p_c=0.6$~GPa.
The high-temperature electronic state can be characterized as a rather poor metallic state with only a small amount of free charge carriers. When cooled down at ambient and weak pressure ($p<p_c$), the system becomes insulating at the quantum Widom line, i.e.\  the spectral weight shifts to higher energies as the Mott gap opens in the optical spectrum. Due to the electron-phonon coupling, we also identify changed in the vibrational features (Fig.~\ref{Fig_Vibrations}).
With $p>p_c$ a metallic ground state is established at reduced temperatures with a pronounced zero-frequency component. When the Mott transition is approached from the metallic side, the increase of correlations strength is monitored quantitatively by the enhancement of the effective mass $m^*$ at $p_c$. Lower-temperature measurements are required for a more detailed characterization of this possibly Fermi-liquid state.

\begin{acknowledgments}
We thank G. Untereiner for preparing the crystals for optical measurements and T. Biesner for the help during the data analysis. This work was partially supported by JSPS KAKENHI grant
No.\ JP16H06346. We acknowledge support by the Deutsche Forschungsgemeinschaft (DFG) via DR228/52-1.
\end{acknowledgments}


\begin{thebibliography}{36}%
\makeatletter
\providecommand \@ifxundefined [1]{%
 \@ifx{#1\undefined}
}%
\providecommand \@ifnum [1]{%
 \ifnum #1\expandafter \@firstoftwo
 \else \expandafter \@secondoftwo
 \fi
}%
\providecommand \@ifx [1]{%
 \ifx #1\expandafter \@firstoftwo
 \else \expandafter \@secondoftwo
 \fi
}%
\providecommand \natexlab [1]{#1}%
\providecommand \enquote  [1]{``#1''}%
\providecommand \bibnamefont  [1]{#1}%
\providecommand \bibfnamefont [1]{#1}%
\providecommand \citenamefont [1]{#1}%
\providecommand \href@noop [0]{\@secondoftwo}%
\providecommand \href [0]{\begingroup \@sanitize@url \@href}%
\providecommand \@href[1]{\@@startlink{#1}\@@href}%
\providecommand \@@href[1]{\endgroup#1\@@endlink}%
\providecommand \@sanitize@url [0]{\catcode `\\12\catcode `\$12\catcode
  `\&12\catcode `\#12\catcode `\^12\catcode `\_12\catcode `\%12\relax}%
\providecommand \@@startlink[1]{}%
\providecommand \@@endlink[0]{}%
\providecommand \url  [0]{\begingroup\@sanitize@url \@url }%
\providecommand \@url [1]{\endgroup\@href {#1}{\urlprefix }}%
\providecommand \urlprefix  [0]{URL }%
\providecommand \Eprint [0]{\href }%
\providecommand \doibase [0]{http://dx.doi.org/}%
\providecommand \selectlanguage [0]{\@gobble}%
\providecommand \bibinfo  [0]{\@secondoftwo}%
\providecommand \bibfield  [0]{\@secondoftwo}%
\providecommand \translation [1]{[#1]}%
\providecommand \BibitemOpen [0]{}%
\providecommand \bibitemStop [0]{}%
\providecommand \bibitemNoStop [0]{.\EOS\space}%
\providecommand \EOS [0]{\spacefactor3000\relax}%
\providecommand \BibitemShut  [1]{\csname bibitem#1\endcsname}%
\let\auto@bib@innerbib\@empty
\bibitem [{\citenamefont {Georges}\ \emph {et~al.}(1996)\citenamefont
  {Georges}, \citenamefont {Kotliar}, \citenamefont {Krauth},\ and\
  \citenamefont {Rozenberg}}]{Georges96}%
  \BibitemOpen
  \bibfield  {author} {\bibinfo {author} {\bibfnamefont {A.}~\bibnamefont
  {Georges}}, \bibinfo {author} {\bibfnamefont {G.}~\bibnamefont {Kotliar}},
  \bibinfo {author} {\bibfnamefont {W.}~\bibnamefont {Krauth}}, \ and\ \bibinfo
  {author} {\bibfnamefont {M.~J.}\ \bibnamefont {Rozenberg}},\ }\href {\doibase
  10.1103/RevModPhys.68.13} {\bibfield  {journal} {\bibinfo  {journal} {Rev.
  Mod. Phys.}\ }\textbf {\bibinfo {volume} {68}},\ \bibinfo {pages} {13}
  (\bibinfo {year} {1996})}\BibitemShut {NoStop}%
\bibitem [{\citenamefont {Kotliar}\ \emph {et~al.}(2006)\citenamefont
  {Kotliar}, \citenamefont {Savrasov}, \citenamefont {Haule}, \citenamefont
  {Oudovenko}, \citenamefont {Parcollet},\ and\ \citenamefont
  {Marianetti}}]{Kotliar06}%
  \BibitemOpen
  \bibfield  {author} {\bibinfo {author} {\bibfnamefont {G.}~\bibnamefont
  {Kotliar}}, \bibinfo {author} {\bibfnamefont {S.~Y.}\ \bibnamefont
  {Savrasov}}, \bibinfo {author} {\bibfnamefont {K.}~\bibnamefont {Haule}},
  \bibinfo {author} {\bibfnamefont {V.~S.}\ \bibnamefont {Oudovenko}}, \bibinfo
  {author} {\bibfnamefont {O.}~\bibnamefont {Parcollet}}, \ and\ \bibinfo
  {author} {\bibfnamefont {C.~A.}\ \bibnamefont {Marianetti}},\ }\href
  {\doibase 10.1103/RevModPhys.78.865} {\bibfield  {journal} {\bibinfo
  {journal} {Rev. Mod. Phys.}\ }\textbf {\bibinfo {volume} {78}},\ \bibinfo
  {pages} {865} (\bibinfo {year} {2006})}\BibitemShut {NoStop}%
\bibitem [{\citenamefont {McWhan}\ \emph {et~al.}(1973)\citenamefont {McWhan},
  \citenamefont {Menth}, \citenamefont {Remeika}, \citenamefont {Brinkman},\
  and\ \citenamefont {Rice}}]{Mcwhan73}%
  \BibitemOpen
  \bibfield  {author} {\bibinfo {author} {\bibfnamefont {D.~B.}\ \bibnamefont
  {McWhan}}, \bibinfo {author} {\bibfnamefont {A.}~\bibnamefont {Menth}},
  \bibinfo {author} {\bibfnamefont {J.~P.}\ \bibnamefont {Remeika}}, \bibinfo
  {author} {\bibfnamefont {W.~F.}\ \bibnamefont {Brinkman}}, \ and\ \bibinfo
  {author} {\bibfnamefont {T.~M.}\ \bibnamefont {Rice}},\ }\href {\doibase
  10.1103/PhysRevB.7.1920} {\bibfield  {journal} {\bibinfo  {journal} {Phys.
  Rev. B}\ }\textbf {\bibinfo {volume} {7}},\ \bibinfo {pages} {1920} (\bibinfo
  {year} {1973})}\BibitemShut {NoStop}%
\bibitem [{\citenamefont {Imada}\ \emph {et~al.}(1998)\citenamefont {Imada},
  \citenamefont {Fujimori},\ and\ \citenamefont {Tokura}}]{Imada98}%
  \BibitemOpen
  \bibfield  {author} {\bibinfo {author} {\bibfnamefont {M.}~\bibnamefont
  {Imada}}, \bibinfo {author} {\bibfnamefont {A.}~\bibnamefont {Fujimori}}, \
  and\ \bibinfo {author} {\bibfnamefont {Y.}~\bibnamefont {Tokura}},\ }\href
  {\doibase 10.1103/RevModPhys.70.1039} {\bibfield  {journal} {\bibinfo
  {journal} {Rev. Mod. Phys.}\ }\textbf {\bibinfo {volume} {70}},\ \bibinfo
  {pages} {1039} (\bibinfo {year} {1998})}\BibitemShut {NoStop}%
\bibitem [{\citenamefont {Limelette}\ \emph {et~al.}(2003)\citenamefont
  {Limelette}, \citenamefont {Georges}, \citenamefont {J{\'e}rome},
  \citenamefont {Wzietek}, \citenamefont {Metcalf},\ and\ \citenamefont
  {Honig}}]{Limelette03}%
  \BibitemOpen
  \bibfield  {author} {\bibinfo {author} {\bibfnamefont {P.}~\bibnamefont
  {Limelette}}, \bibinfo {author} {\bibfnamefont {A.}~\bibnamefont {Georges}},
  \bibinfo {author} {\bibfnamefont {D.}~\bibnamefont {J{\'e}rome}}, \bibinfo
  {author} {\bibfnamefont {P.}~\bibnamefont {Wzietek}}, \bibinfo {author}
  {\bibfnamefont {P.}~\bibnamefont {Metcalf}}, \ and\ \bibinfo {author}
  {\bibfnamefont {J.}~\bibnamefont {Honig}},\ }\href@noop {} {\bibfield
  {journal} {\bibinfo  {journal} {Science}\ }\textbf {\bibinfo {volume}
  {302}},\ \bibinfo {pages} {89} (\bibinfo {year} {2003})}\BibitemShut
  {NoStop}%
\bibitem [{\citenamefont {Arcangeletti}\ \emph {et~al.}(2007)\citenamefont
  {Arcangeletti}, \citenamefont {Baldassarre}, \citenamefont {Di~Castro},
  \citenamefont {Lupi}, \citenamefont {Malavasi}, \citenamefont {Marini},
  \citenamefont {Perucchi},\ and\ \citenamefont {Postorino}}]{Arcangeletti07}%
  \BibitemOpen
  \bibfield  {author} {\bibinfo {author} {\bibfnamefont {E.}~\bibnamefont
  {Arcangeletti}}, \bibinfo {author} {\bibfnamefont {L.}~\bibnamefont
  {Baldassarre}}, \bibinfo {author} {\bibfnamefont {D.}~\bibnamefont
  {Di~Castro}}, \bibinfo {author} {\bibfnamefont {S.}~\bibnamefont {Lupi}},
  \bibinfo {author} {\bibfnamefont {L.}~\bibnamefont {Malavasi}}, \bibinfo
  {author} {\bibfnamefont {C.}~\bibnamefont {Marini}}, \bibinfo {author}
  {\bibfnamefont {A.}~\bibnamefont {Perucchi}}, \ and\ \bibinfo {author}
  {\bibfnamefont {P.}~\bibnamefont {Postorino}},\ }\href {\doibase
  10.1103/PhysRevLett.98.196406} {\bibfield  {journal} {\bibinfo  {journal}
  {Phys. Rev. Lett.}\ }\textbf {\bibinfo {volume} {98}},\ \bibinfo {pages}
  {196406} (\bibinfo {year} {2007})}\BibitemShut {NoStop}%
\bibitem [{\citenamefont {Norman}(2016)}]{Norman16}%
  \BibitemOpen
  \bibfield  {author} {\bibinfo {author} {\bibfnamefont {M.~R.}\ \bibnamefont
  {Norman}},\ }\href {\doibase 10.1103/RevModPhys.88.041002} {\bibfield
  {journal} {\bibinfo  {journal} {Rev. Mod. Phys.}\ }\textbf {\bibinfo {volume}
  {88}},\ \bibinfo {pages} {041002} (\bibinfo {year} {2016})}\BibitemShut
  {NoStop}%
\bibitem [{\citenamefont {Savary}\ and\ \citenamefont
  {Balents}(2017)}]{Savary17}%
  \BibitemOpen
  \bibfield  {author} {\bibinfo {author} {\bibfnamefont {L.}~\bibnamefont
  {Savary}}\ and\ \bibinfo {author} {\bibfnamefont {L.}~\bibnamefont
  {Balents}},\ }\href {\doibase 10.1088/0034-4885/80/1/016502} {\bibfield
  {journal} {\bibinfo  {journal} {Rep. Progr. Phys.}\ }\textbf {\bibinfo
  {volume} {80}},\ \bibinfo {pages} {016502} (\bibinfo {year}
  {2017})}\BibitemShut {NoStop}%
\bibitem [{\citenamefont {Zhou}\ \emph {et~al.}(2017)\citenamefont {Zhou},
  \citenamefont {Kanoda},\ and\ \citenamefont {Ng}}]{Zhou17}%
  \BibitemOpen
  \bibfield  {author} {\bibinfo {author} {\bibfnamefont {Y.}~\bibnamefont
  {Zhou}}, \bibinfo {author} {\bibfnamefont {K.}~\bibnamefont {Kanoda}}, \ and\
  \bibinfo {author} {\bibfnamefont {T.-K.}\ \bibnamefont {Ng}},\ }\href
  {https://link.aps.org/doi/10.1103/RevModPhys.89.025003} {\bibfield  {journal}
  {\bibinfo  {journal} {Rev. Mod. Phys.}\ }\textbf {\bibinfo {volume} {89}},\
  \bibinfo {pages} {25003} (\bibinfo {year} {2017})}\BibitemShut {NoStop}%
\bibitem [{\citenamefont {Dressel}\ and\ \citenamefont
  {Pustogow}(2018)}]{Dressel18}%
  \BibitemOpen
  \bibfield  {author} {\bibinfo {author} {\bibfnamefont {M.}~\bibnamefont
  {Dressel}}\ and\ \bibinfo {author} {\bibfnamefont {A.}~\bibnamefont
  {Pustogow}},\ }\href {http://stacks.iop.org/0953-8984/30/i=20/a=203001}
  {\bibfield  {journal} {\bibinfo  {journal} {J. Phys.: Condens. Matter}\
  }\textbf {\bibinfo {volume} {30}},\ \bibinfo {pages} {203001} (\bibinfo
  {year} {2018})}\BibitemShut {NoStop}%
\bibitem [{\citenamefont {Kanoda}\ and\ \citenamefont {Kato}(2011)}]{Kanoda11}%
  \BibitemOpen
  \bibfield  {author} {\bibinfo {author} {\bibfnamefont {K.}~\bibnamefont
  {Kanoda}}\ and\ \bibinfo {author} {\bibfnamefont {R.}~\bibnamefont {Kato}},\
  }\href {\doibase 10.1146/annurev-conmatphys-062910-140521} {\bibfield
  {journal} {\bibinfo  {journal} {Annu. Rev. Condens. Matter Phys.}\ }\textbf
  {\bibinfo {volume} {2}},\ \bibinfo {pages} {167} (\bibinfo {year}
  {2011})}\BibitemShut {NoStop}%
\bibitem [{\citenamefont {Pustogow}\ \emph {et~al.}(2018)\citenamefont
  {Pustogow}, \citenamefont {Bories}, \citenamefont {L{\"o}hle}, \citenamefont
  {R{\"o}sslhuber}, \citenamefont {Zhukova}, \citenamefont {Gorshunov},
  \citenamefont {Tomi{\'c}}, \citenamefont {Schlueter}, \citenamefont
  {H{\"u}bner}, \citenamefont {Hiramatsu}, \citenamefont {Yoshida},
  \citenamefont {Saito}, \citenamefont {Kato}, \citenamefont {Lee},
  \citenamefont {Dobrosavljevi{\'c}}, \citenamefont {Fratini},\ and\
  \citenamefont {Dressel}}]{Pustogow17}%
  \BibitemOpen
  \bibfield  {author} {\bibinfo {author} {\bibfnamefont {A.}~\bibnamefont
  {Pustogow}}, \bibinfo {author} {\bibfnamefont {M.}~\bibnamefont {Bories}},
  \bibinfo {author} {\bibfnamefont {A.}~\bibnamefont {L{\"o}hle}}, \bibinfo
  {author} {\bibfnamefont {R.}~\bibnamefont {R{\"o}sslhuber}}, \bibinfo
  {author} {\bibfnamefont {E.}~\bibnamefont {Zhukova}}, \bibinfo {author}
  {\bibfnamefont {B.}~\bibnamefont {Gorshunov}}, \bibinfo {author}
  {\bibfnamefont {S.}~\bibnamefont {Tomi{\'c}}}, \bibinfo {author}
  {\bibfnamefont {J.}~\bibnamefont {Schlueter}}, \bibinfo {author}
  {\bibfnamefont {R.}~\bibnamefont {H{\"u}bner}}, \bibinfo {author}
  {\bibfnamefont {T.}~\bibnamefont {Hiramatsu}}, \bibinfo {author}
  {\bibfnamefont {Y.}~\bibnamefont {Yoshida}}, \bibinfo {author} {\bibfnamefont
  {G.}~\bibnamefont {Saito}}, \bibinfo {author} {\bibfnamefont
  {R.}~\bibnamefont {Kato}}, \bibinfo {author} {\bibfnamefont {T.-H.}\
  \bibnamefont {Lee}}, \bibinfo {author} {\bibfnamefont {V.}~\bibnamefont
  {Dobrosavljevi{\'c}}}, \bibinfo {author} {\bibfnamefont {S.}~\bibnamefont
  {Fratini}}, \ and\ \bibinfo {author} {\bibfnamefont {M.}~\bibnamefont
  {Dressel}},\ }\href@noop {} {\bibfield  {journal} {\bibinfo  {journal} {Nat.
  Mater.}\ }\textbf {\bibinfo {volume} {17}},\ \bibinfo {pages} {773} (\bibinfo
  {year} {2018})}\BibitemShut {NoStop}%
\bibitem [{\citenamefont {Itou}\ \emph {et~al.}(2017)\citenamefont {Itou},
  \citenamefont {Watanabe}, \citenamefont {Maegawa}, \citenamefont {Tajima},
  \citenamefont {Tajima}, \citenamefont {Kubo}, \citenamefont {Kato},\ and\
  \citenamefont {Kanoda}}]{Itou17}%
  \BibitemOpen
  \bibfield  {author} {\bibinfo {author} {\bibfnamefont {T.}~\bibnamefont
  {Itou}}, \bibinfo {author} {\bibfnamefont {E.}~\bibnamefont {Watanabe}},
  \bibinfo {author} {\bibfnamefont {S.}~\bibnamefont {Maegawa}}, \bibinfo
  {author} {\bibfnamefont {A.}~\bibnamefont {Tajima}}, \bibinfo {author}
  {\bibfnamefont {N.}~\bibnamefont {Tajima}}, \bibinfo {author} {\bibfnamefont
  {K.}~\bibnamefont {Kubo}}, \bibinfo {author} {\bibfnamefont {R.}~\bibnamefont
  {Kato}}, \ and\ \bibinfo {author} {\bibfnamefont {K.}~\bibnamefont
  {Kanoda}},\ }\href {\doibase 10.1126/sciadv.1601594} {\bibfield  {journal}
  {\bibinfo  {journal} {Sci. Adv.}\ }\textbf {\bibinfo {volume} {3}},\ \bibinfo
  {pages} {e1601594} (\bibinfo {year} {2017})}\BibitemShut {NoStop}%
\bibitem [{\citenamefont {Kato}\ \emph {et~al.}(2012)\citenamefont {Kato},
  \citenamefont {Fukunaga}, \citenamefont {Yamamoto}, \citenamefont {Ueda},\
  and\ \citenamefont {Hengbo}}]{Kato12}%
  \BibitemOpen
  \bibfield  {author} {\bibinfo {author} {\bibfnamefont {R.}~\bibnamefont
  {Kato}}, \bibinfo {author} {\bibfnamefont {T.}~\bibnamefont {Fukunaga}},
  \bibinfo {author} {\bibfnamefont {H.~M.}\ \bibnamefont {Yamamoto}}, \bibinfo
  {author} {\bibfnamefont {K.}~\bibnamefont {Ueda}}, \ and\ \bibinfo {author}
  {\bibfnamefont {C.}~\bibnamefont {Hengbo}},\ }\href {\doibase
  10.1002/pssb.201100645} {\bibfield  {journal} {\bibinfo  {journal} {phys.
  stat. sol. (b)}\ }\textbf {\bibinfo {volume} {249}},\ \bibinfo {pages} {999}
  (\bibinfo {year} {2012})}\BibitemShut {NoStop}%
\bibitem [{\citenamefont {Scriven}\ and\ \citenamefont
  {Powell}(2012)}]{Scriven12}%
  \BibitemOpen
  \bibfield  {author} {\bibinfo {author} {\bibfnamefont {E.~P.}\ \bibnamefont
  {Scriven}}\ and\ \bibinfo {author} {\bibfnamefont {B.~J.}\ \bibnamefont
  {Powell}},\ }\href {\doibase 10.1103/PhysRevLett.109.097206} {\bibfield
  {journal} {\bibinfo  {journal} {Phys. Rev. Lett.}\ }\textbf {\bibinfo
  {volume} {109}},\ \bibinfo {pages} {097206} (\bibinfo {year}
  {2012})}\BibitemShut {NoStop}%
\bibitem [{\citenamefont {Nakamura}\ \emph {et~al.}(2012)\citenamefont
  {Nakamura}, \citenamefont {Yoshimoto},\ and\ \citenamefont
  {Imada}}]{Nakamura12}%
  \BibitemOpen
  \bibfield  {author} {\bibinfo {author} {\bibfnamefont {K.}~\bibnamefont
  {Nakamura}}, \bibinfo {author} {\bibfnamefont {Y.}~\bibnamefont {Yoshimoto}},
  \ and\ \bibinfo {author} {\bibfnamefont {M.}~\bibnamefont {Imada}},\ }\href
  {\doibase 10.1103/PhysRevB.86.205117} {\bibfield  {journal} {\bibinfo
  {journal} {Phys. Rev. B}\ }\textbf {\bibinfo {volume} {86}},\ \bibinfo
  {pages} {205117} (\bibinfo {year} {2012})}\BibitemShut {NoStop}%
\bibitem [{\citenamefont {Ueda}\ \emph {et~al.}(2018)\citenamefont {Ueda},
  \citenamefont {Tsumuraya},\ and\ \citenamefont {Kato}}]{Ueda18}%
  \BibitemOpen
  \bibfield  {author} {\bibinfo {author} {\bibfnamefont {K.}~\bibnamefont
  {Ueda}}, \bibinfo {author} {\bibfnamefont {T.}~\bibnamefont {Tsumuraya}}, \
  and\ \bibinfo {author} {\bibfnamefont {R.}~\bibnamefont {Kato}},\ }\href@noop
  {} {\bibfield  {journal} {\bibinfo  {journal} {Crystals}\ }\textbf {\bibinfo
  {volume} {8}} (\bibinfo {year} {2018})}\BibitemShut {NoStop}%
\bibitem [{\citenamefont {Beyer}\ and\ \citenamefont
  {Dressel}(2015)}]{Beyer15}%
  \BibitemOpen
  \bibfield  {author} {\bibinfo {author} {\bibfnamefont {R.}~\bibnamefont
  {Beyer}}\ and\ \bibinfo {author} {\bibfnamefont {M.}~\bibnamefont
  {Dressel}},\ }\href {\doibase 10.1063/1.4920921} {\bibfield  {journal}
  {\bibinfo  {journal} {Rev. Sci. Instr.}\ }\textbf {\bibinfo {volume} {86}},\
  \bibinfo {pages} {053904} (\bibinfo {year} {2015})}\BibitemShut {NoStop}%
\bibitem [{\citenamefont {Dressel}\ and\ \citenamefont
  {Gr{\"u}ner}(2002)}]{DresselGruner02}%
  \BibitemOpen
  \bibfield  {author} {\bibinfo {author} {\bibfnamefont {M.}~\bibnamefont
  {Dressel}}\ and\ \bibinfo {author} {\bibfnamefont {G.}~\bibnamefont
  {Gr{\"u}ner}},\ }\href@noop {} {\emph {\bibinfo {title} {Electrodynamics of
  Solids}}}\ (\bibinfo  {publisher} {Cambridge University Press},\ \bibinfo
  {address} {Cambridge},\ \bibinfo {year} {2002})\BibitemShut {NoStop}%
\bibitem [{SM()}]{SM}%
  \BibitemOpen
  \href@noop {} {}\bibinfo {note} {In the Supplemental Material at
  http://link.aps.org./supplemental/ we plot the raw data of the measured
  reflectivity spectra at all temperatures, pressure values and both
  polarizations, $E\parallel a$ and $E\parallel b$. Also shown is the
  low-temperature conductivity at different pressure values together with the
  corresponding Drude-Lorentz fits used to extract the Coulomb correlation $U$
  and bandwidth $W$ in the insulating cases. For the metallic cases, the
  spectral weight of the Drude and intraband contributions are
  shown.}\BibitemShut {Stop}%
\bibitem [{\citenamefont {Dressel}\ and\ \citenamefont
  {Drichko}(2004)}]{Dressel04}%
  \BibitemOpen
  \bibfield  {author} {\bibinfo {author} {\bibfnamefont {M.}~\bibnamefont
  {Dressel}}\ and\ \bibinfo {author} {\bibfnamefont {N.}~\bibnamefont
  {Drichko}},\ }\href {\doibase 10.1021/cr030642f} {\bibfield  {journal}
  {\bibinfo  {journal} {Chem. Rev.}\ }\textbf {\bibinfo {volume} {104}},\
  \bibinfo {pages} {5689} (\bibinfo {year} {2004})}\BibitemShut {NoStop}%
\bibitem [{\citenamefont {Yamamoto}\ \emph
  {et~al.}(2011{\natexlab{a}})\citenamefont {Yamamoto}, \citenamefont
  {Nakazawa}, \citenamefont {Tamura}, \citenamefont {Fukunaga}, \citenamefont
  {Kato},\ and\ \citenamefont {Yakushi}}]{Yamamoto11a}%
  \BibitemOpen
  \bibfield  {author} {\bibinfo {author} {\bibfnamefont {T.}~\bibnamefont
  {Yamamoto}}, \bibinfo {author} {\bibfnamefont {Y.}~\bibnamefont {Nakazawa}},
  \bibinfo {author} {\bibfnamefont {M.}~\bibnamefont {Tamura}}, \bibinfo
  {author} {\bibfnamefont {T.}~\bibnamefont {Fukunaga}}, \bibinfo {author}
  {\bibfnamefont {R.}~\bibnamefont {Kato}}, \ and\ \bibinfo {author}
  {\bibfnamefont {K.}~\bibnamefont {Yakushi}},\ }\href {\doibase
  10.1143/JPSJ.80.074717} {\bibfield  {journal} {\bibinfo  {journal} {J. Phys.
  Soc. Jpn.}\ }\textbf {\bibinfo {volume} {80}},\ \bibinfo {pages} {074717}
  (\bibinfo {year} {2011}{\natexlab{a}})}\BibitemShut {NoStop}%
\bibitem [{\citenamefont {Yamamoto}\ \emph
  {et~al.}(2011{\natexlab{b}})\citenamefont {Yamamoto}, \citenamefont
  {Nakazawa}, \citenamefont {Tamura}, \citenamefont {Nakao}, \citenamefont
  {Ikemoto}, \citenamefont {Moriwaki}, \citenamefont {Fukaya}, \citenamefont
  {Kato},\ and\ \citenamefont {Yakushi}}]{Yamamoto11b}%
  \BibitemOpen
  \bibfield  {author} {\bibinfo {author} {\bibfnamefont {T.}~\bibnamefont
  {Yamamoto}}, \bibinfo {author} {\bibfnamefont {Y.}~\bibnamefont {Nakazawa}},
  \bibinfo {author} {\bibfnamefont {M.}~\bibnamefont {Tamura}}, \bibinfo
  {author} {\bibfnamefont {A.}~\bibnamefont {Nakao}}, \bibinfo {author}
  {\bibfnamefont {Y.}~\bibnamefont {Ikemoto}}, \bibinfo {author} {\bibfnamefont
  {T.}~\bibnamefont {Moriwaki}}, \bibinfo {author} {\bibfnamefont
  {A.}~\bibnamefont {Fukaya}}, \bibinfo {author} {\bibfnamefont
  {R.}~\bibnamefont {Kato}}, \ and\ \bibinfo {author} {\bibfnamefont
  {K.}~\bibnamefont {Yakushi}},\ }\href {\doibase 10.1143/JPSJ.80.123709}
  {\bibfield  {journal} {\bibinfo  {journal} {J. Phys. Soc. Jpn.}\ }\textbf
  {\bibinfo {volume} {80}},\ \bibinfo {pages} {123709} (\bibinfo {year}
  {2011}{\natexlab{b}})}\BibitemShut {NoStop}%
\bibitem [{\citenamefont {Yamamoto}\ \emph {et~al.}(2017)\citenamefont
  {Yamamoto}, \citenamefont {Fujimoto}, \citenamefont {Naito}, \citenamefont
  {Nakazawa}, \citenamefont {Tamura}, \citenamefont {Yakushi}, \citenamefont
  {Ikemoto}, \citenamefont {Moriwaki},\ and\ \citenamefont
  {Kato}}]{Yamamoto17}%
  \BibitemOpen
  \bibfield  {author} {\bibinfo {author} {\bibfnamefont {T.}~\bibnamefont
  {Yamamoto}}, \bibinfo {author} {\bibfnamefont {T.}~\bibnamefont {Fujimoto}},
  \bibinfo {author} {\bibfnamefont {T.}~\bibnamefont {Naito}}, \bibinfo
  {author} {\bibfnamefont {Y.}~\bibnamefont {Nakazawa}}, \bibinfo {author}
  {\bibfnamefont {M.}~\bibnamefont {Tamura}}, \bibinfo {author} {\bibfnamefont
  {K.}~\bibnamefont {Yakushi}}, \bibinfo {author} {\bibfnamefont
  {Y.}~\bibnamefont {Ikemoto}}, \bibinfo {author} {\bibfnamefont
  {T.}~\bibnamefont {Moriwaki}}, \ and\ \bibinfo {author} {\bibfnamefont
  {R.}~\bibnamefont {Kato}},\ }\href@noop {} {\bibfield  {journal} {\bibinfo
  {journal} {Sci. Rep.}\ }\textbf {\bibinfo {volume} {7}},\ \bibinfo {pages}
  {12930} (\bibinfo {year} {2017})}\BibitemShut {NoStop}%
\bibitem [{\citenamefont {Liu}\ \emph {et~al.}(2016)\citenamefont {Liu},
  \citenamefont {Ishikawa}, \citenamefont {Takehara}, \citenamefont {Miyagawa},
  \citenamefont {Tamura},\ and\ \citenamefont {Kanoda}}]{Liu16}%
  \BibitemOpen
  \bibfield  {author} {\bibinfo {author} {\bibfnamefont {D.}~\bibnamefont
  {Liu}}, \bibinfo {author} {\bibfnamefont {K.}~\bibnamefont {Ishikawa}},
  \bibinfo {author} {\bibfnamefont {R.}~\bibnamefont {Takehara}}, \bibinfo
  {author} {\bibfnamefont {K.}~\bibnamefont {Miyagawa}}, \bibinfo {author}
  {\bibfnamefont {M.}~\bibnamefont {Tamura}}, \ and\ \bibinfo {author}
  {\bibfnamefont {K.}~\bibnamefont {Kanoda}},\ }\href {\doibase
  10.1103/PhysRevLett.116.226401} {\bibfield  {journal} {\bibinfo  {journal}
  {Phys. Rev. Lett.}\ }\textbf {\bibinfo {volume} {116}},\ \bibinfo {pages}
  {226401} (\bibinfo {year} {2016})}\BibitemShut {NoStop}%
\bibitem [{\citenamefont {Uykur}\ \emph {et~al.}(2018)\citenamefont {Uykur},
  \citenamefont {Li}, \citenamefont {Kuntscher},\ and\ \citenamefont
  {Dressel}}]{Uykur18}%
  \BibitemOpen
  \bibfield  {author} {\bibinfo {author} {\bibfnamefont {E.}~\bibnamefont
  {Uykur}}, \bibinfo {author} {\bibfnamefont {W.}~\bibnamefont {Li}}, \bibinfo
  {author} {\bibfnamefont {C.~A.}\ \bibnamefont {Kuntscher}}, \ and\ \bibinfo
  {author} {\bibfnamefont {M.}~\bibnamefont {Dressel}},\ }\href@noop {}
  {\bibfield  {journal} {\bibinfo  {journal} {arXiv:1803.00755}\ } (\bibinfo
  {year} {2018})}\BibitemShut {NoStop}%
\bibitem [{\citenamefont {Basov}\ \emph {et~al.}(2011)\citenamefont {Basov},
  \citenamefont {Averitt}, \citenamefont {van~der Marel}, \citenamefont
  {Dressel},\ and\ \citenamefont {Haule}}]{Basov11}%
  \BibitemOpen
  \bibfield  {author} {\bibinfo {author} {\bibfnamefont {D.~N.}\ \bibnamefont
  {Basov}}, \bibinfo {author} {\bibfnamefont {R.~D.}\ \bibnamefont {Averitt}},
  \bibinfo {author} {\bibfnamefont {D.}~\bibnamefont {van~der Marel}}, \bibinfo
  {author} {\bibfnamefont {M.}~\bibnamefont {Dressel}}, \ and\ \bibinfo
  {author} {\bibfnamefont {K.}~\bibnamefont {Haule}},\ }\href {\doibase
  10.1103/RevModPhys.83.471} {\bibfield  {journal} {\bibinfo  {journal} {Rev.
  Mod. Phys.}\ }\textbf {\bibinfo {volume} {83}},\ \bibinfo {pages} {471}
  (\bibinfo {year} {2011})}\BibitemShut {NoStop}%
\bibitem [{\citenamefont {Merino}\ \emph {et~al.}(2008)\citenamefont {Merino},
  \citenamefont {Dumm}, \citenamefont {Drichko}, \citenamefont {Dressel},\ and\
  \citenamefont {McKenzie}}]{Merino08}%
  \BibitemOpen
  \bibfield  {author} {\bibinfo {author} {\bibfnamefont {J.}~\bibnamefont
  {Merino}}, \bibinfo {author} {\bibfnamefont {M.}~\bibnamefont {Dumm}},
  \bibinfo {author} {\bibfnamefont {N.}~\bibnamefont {Drichko}}, \bibinfo
  {author} {\bibfnamefont {M.}~\bibnamefont {Dressel}}, \ and\ \bibinfo
  {author} {\bibfnamefont {R.~H.}\ \bibnamefont {McKenzie}},\ }\href {\doibase
  10.1103/PhysRevLett.100.086404} {\bibfield  {journal} {\bibinfo  {journal}
  {Phys. Rev. Lett.}\ }\textbf {\bibinfo {volume} {100}},\ \bibinfo {pages}
  {086404} (\bibinfo {year} {2008})}\BibitemShut {NoStop}%
\bibitem [{\citenamefont {Rozenberg}\ \emph {et~al.}(1995)\citenamefont
  {Rozenberg}, \citenamefont {Kotliar}, \citenamefont {Kajueter}, \citenamefont
  {Thomas}, \citenamefont {Rapkine}, \citenamefont {Honig},\ and\ \citenamefont
  {Metcalf}}]{Rozenberg95}%
  \BibitemOpen
  \bibfield  {author} {\bibinfo {author} {\bibfnamefont {M.~J.}\ \bibnamefont
  {Rozenberg}}, \bibinfo {author} {\bibfnamefont {G.}~\bibnamefont {Kotliar}},
  \bibinfo {author} {\bibfnamefont {H.}~\bibnamefont {Kajueter}}, \bibinfo
  {author} {\bibfnamefont {G.~A.}\ \bibnamefont {Thomas}}, \bibinfo {author}
  {\bibfnamefont {D.~H.}\ \bibnamefont {Rapkine}}, \bibinfo {author}
  {\bibfnamefont {J.~M.}\ \bibnamefont {Honig}}, \ and\ \bibinfo {author}
  {\bibfnamefont {P.}~\bibnamefont {Metcalf}},\ }\href@noop {} {\bibfield
  {journal} {\bibinfo  {journal} {Phys. Rev. Lett.}\ }\textbf {\bibinfo
  {volume} {75}},\ \bibinfo {pages} {105} (\bibinfo {year} {1995})}\BibitemShut
  {NoStop}%
\bibitem [{\citenamefont {Vu\ifmmode \check{c}\else \v{c}\fi{}i\ifmmode
  \check{c}\else \v{c}\fi{}evi\ifmmode~\acute{c}\else \'{c}\fi{}}\ \emph
  {et~al.}(2013)\citenamefont {Vu\ifmmode \check{c}\else \v{c}\fi{}i\ifmmode
  \check{c}\else \v{c}\fi{}evi\ifmmode~\acute{c}\else \'{c}\fi{}},
  \citenamefont {Terletska}, \citenamefont {Tanaskovi\ifmmode~\acute{c}\else
  \'{c}\fi{}},\ and\ \citenamefont {Dobrosavljevi\ifmmode~\acute{c}\else
  \'{c}\fi{}}}]{Vucicevic13}%
  \BibitemOpen
  \bibfield  {author} {\bibinfo {author} {\bibfnamefont {J.}~\bibnamefont
  {Vu\ifmmode \check{c}\else \v{c}\fi{}i\ifmmode \check{c}\else
  \v{c}\fi{}evi\ifmmode~\acute{c}\else \'{c}\fi{}}}, \bibinfo {author}
  {\bibfnamefont {H.}~\bibnamefont {Terletska}}, \bibinfo {author}
  {\bibfnamefont {D.}~\bibnamefont {Tanaskovi\ifmmode~\acute{c}\else
  \'{c}\fi{}}}, \ and\ \bibinfo {author} {\bibfnamefont {V.}~\bibnamefont
  {Dobrosavljevi\ifmmode~\acute{c}\else \'{c}\fi{}}},\ }\href {\doibase
  10.1103/PhysRevB.88.075143} {\bibfield  {journal} {\bibinfo  {journal} {Phys.
  Rev. B}\ }\textbf {\bibinfo {volume} {88}},\ \bibinfo {pages} {075143}
  (\bibinfo {year} {2013})}\BibitemShut {NoStop}%
\bibitem [{\citenamefont {Klehe}\ \emph {et~al.}(2000)\citenamefont {Klehe},
  \citenamefont {McDonald}, \citenamefont {Goncharov}, \citenamefont
  {Struzhkin}, \citenamefont {kwang Mao}, \citenamefont {Hemley}, \citenamefont
  {Sasaki}, \citenamefont {Hayes},\ and\ \citenamefont {Singleton}}]{Klehe00}%
  \BibitemOpen
  \bibfield  {author} {\bibinfo {author} {\bibfnamefont {A.-K.}\ \bibnamefont
  {Klehe}}, \bibinfo {author} {\bibfnamefont {R.~D.}\ \bibnamefont {McDonald}},
  \bibinfo {author} {\bibfnamefont {A.~F.}\ \bibnamefont {Goncharov}}, \bibinfo
  {author} {\bibfnamefont {V.~V.}\ \bibnamefont {Struzhkin}}, \bibinfo {author}
  {\bibfnamefont {H.}~\bibnamefont {kwang Mao}}, \bibinfo {author}
  {\bibfnamefont {R.~J.}\ \bibnamefont {Hemley}}, \bibinfo {author}
  {\bibfnamefont {T.}~\bibnamefont {Sasaki}}, \bibinfo {author} {\bibfnamefont
  {W.}~\bibnamefont {Hayes}}, \ and\ \bibinfo {author} {\bibfnamefont
  {J.}~\bibnamefont {Singleton}},\ }\href
  {http://stacks.iop.org/0953-8984/12/i=13/a=103} {\bibfield  {journal}
  {\bibinfo  {journal} {J. Phys.: Condens. Matter}\ }\textbf {\bibinfo {volume}
  {12}},\ \bibinfo {pages} {L247} (\bibinfo {year} {2000})}\BibitemShut
  {NoStop}%
\bibitem [{\citenamefont {McDonald}\ \emph {et~al.}(2003)\citenamefont
  {McDonald}, \citenamefont {Klehe}, \citenamefont {Singleton},\ and\
  \citenamefont {Hayes}}]{McDonald03}%
  \BibitemOpen
  \bibfield  {author} {\bibinfo {author} {\bibfnamefont {R.}~\bibnamefont
  {McDonald}}, \bibinfo {author} {\bibfnamefont {A.}~\bibnamefont {Klehe}},
  \bibinfo {author} {\bibfnamefont {J.}~\bibnamefont {Singleton}}, \ and\
  \bibinfo {author} {\bibfnamefont {W.}~\bibnamefont {Hayes}},\ }\href@noop {}
  {\bibfield  {journal} {\bibinfo  {journal} {J. Phys.: Condens. Matter}\
  }\textbf {\bibinfo {volume} {15}},\ \bibinfo {pages} {5315} (\bibinfo {year}
  {2003})}\BibitemShut {NoStop}%
\bibitem [{\citenamefont {Dumm}\ \emph {et~al.}(2009)\citenamefont {Dumm},
  \citenamefont {Faltermeier}, \citenamefont {Drichko}, \citenamefont
  {Dressel}, \citenamefont {M\'ezi\`ere},\ and\ \citenamefont
  {Batail}}]{Dumm09}%
  \BibitemOpen
  \bibfield  {author} {\bibinfo {author} {\bibfnamefont {M.}~\bibnamefont
  {Dumm}}, \bibinfo {author} {\bibfnamefont {D.}~\bibnamefont {Faltermeier}},
  \bibinfo {author} {\bibfnamefont {N.}~\bibnamefont {Drichko}}, \bibinfo
  {author} {\bibfnamefont {M.}~\bibnamefont {Dressel}}, \bibinfo {author}
  {\bibfnamefont {C.}~\bibnamefont {M\'ezi\`ere}}, \ and\ \bibinfo {author}
  {\bibfnamefont {P.}~\bibnamefont {Batail}},\ }\href {\doibase
  10.1103/PhysRevB.79.195106} {\bibfield  {journal} {\bibinfo  {journal} {Phys.
  Rev. B}\ }\textbf {\bibinfo {volume} {79}},\ \bibinfo {pages} {195106}
  (\bibinfo {year} {2009})}\BibitemShut {NoStop}%
\bibitem [{\citenamefont {Gati}\ \emph {et~al.}(2016)\citenamefont {Gati},
  \citenamefont {Garst}, \citenamefont {Manna}, \citenamefont {Tutsch},
  \citenamefont {Wolf}, \citenamefont {Bartosch}, \citenamefont {Schubert},
  \citenamefont {Sasaki}, \citenamefont {Schlueter},\ and\ \citenamefont
  {Lang}}]{Gati16}%
  \BibitemOpen
  \bibfield  {author} {\bibinfo {author} {\bibfnamefont {E.}~\bibnamefont
  {Gati}}, \bibinfo {author} {\bibfnamefont {M.}~\bibnamefont {Garst}},
  \bibinfo {author} {\bibfnamefont {R.~S.}\ \bibnamefont {Manna}}, \bibinfo
  {author} {\bibfnamefont {U.}~\bibnamefont {Tutsch}}, \bibinfo {author}
  {\bibfnamefont {B.}~\bibnamefont {Wolf}}, \bibinfo {author} {\bibfnamefont
  {L.}~\bibnamefont {Bartosch}}, \bibinfo {author} {\bibfnamefont
  {H.}~\bibnamefont {Schubert}}, \bibinfo {author} {\bibfnamefont
  {T.}~\bibnamefont {Sasaki}}, \bibinfo {author} {\bibfnamefont {J.~A.}\
  \bibnamefont {Schlueter}}, \ and\ \bibinfo {author} {\bibfnamefont
  {M.}~\bibnamefont {Lang}},\ }\href {\doibase 10.1126/sciadv.1601646}
  {\bibfield  {journal} {\bibinfo  {journal} {Sci. Adv.}\ }\textbf {\bibinfo
  {volume} {2}},\ \bibinfo {pages} {e1601646} (\bibinfo {year}
  {2016})}\BibitemShut {NoStop}%
\bibitem [{\citenamefont {Kawasugi}\ \emph {et~al.}(2016)\citenamefont
  {Kawasugi}, \citenamefont {Seki}, \citenamefont {Edagawa}, \citenamefont
  {Sato}, \citenamefont {Pu}, \citenamefont {Takenobu}, \citenamefont {Yunoki},
  \citenamefont {Yamamoto},\ and\ \citenamefont {Kato}}]{Kawasugi16}%
  \BibitemOpen
  \bibfield  {author} {\bibinfo {author} {\bibfnamefont {Y.}~\bibnamefont
  {Kawasugi}}, \bibinfo {author} {\bibfnamefont {K.}~\bibnamefont {Seki}},
  \bibinfo {author} {\bibfnamefont {Y.}~\bibnamefont {Edagawa}}, \bibinfo
  {author} {\bibfnamefont {Y.}~\bibnamefont {Sato}}, \bibinfo {author}
  {\bibfnamefont {J.}~\bibnamefont {Pu}}, \bibinfo {author} {\bibfnamefont
  {T.}~\bibnamefont {Takenobu}}, \bibinfo {author} {\bibfnamefont
  {S.}~\bibnamefont {Yunoki}}, \bibinfo {author} {\bibfnamefont {H.~M.}\
  \bibnamefont {Yamamoto}}, \ and\ \bibinfo {author} {\bibfnamefont
  {R.}~\bibnamefont {Kato}},\ }\href {\doibase 10.1038/ncomms12356} {\bibfield
  {journal} {\bibinfo  {journal} {Nature Commun.}\ }\textbf {\bibinfo {volume}
  {7}},\ \bibinfo {pages} {12356} (\bibinfo {year} {2016})}\BibitemShut
  {NoStop}%
\bibitem [{\citenamefont {\ifmmode~\check{C}\else \v{C}\fi{}ulo}\ \emph
  {et~al.}(2019)\citenamefont {\ifmmode~\check{C}\else \v{C}\fi{}ulo},
  \citenamefont {Tafra}, \citenamefont {Mihaljevi\ifmmode~\acute{c}\else
  \'{c}\fi{}}, \citenamefont {Basleti\ifmmode~\acute{c}\else \'{c}\fi{}},
  \citenamefont {Kuve\ifmmode \check{z}\else
  \v{z}\fi{}di\ifmmode~\acute{c}\else \'{c}\fi{}}, \citenamefont {Ivek},
  \citenamefont {Hamzi\ifmmode~\acute{c}\else \'{c}\fi{}}, \citenamefont
  {Tomi\ifmmode~\acute{c}\else \'{c}\fi{}}, \citenamefont {Hiramatsu},
  \citenamefont {Yoshida}, \citenamefont {Saito}, \citenamefont {Schlueter},
  \citenamefont {Dressel},\ and\ \citenamefont
  {Korin-Hamzi\ifmmode~\acute{c}\else \'{c}\fi{}}}]{Culo19}%
  \BibitemOpen
  \bibfield  {author} {\bibinfo {author} {\bibfnamefont {M.}~\bibnamefont
  {\ifmmode~\check{C}\else \v{C}\fi{}ulo}}, \bibinfo {author} {\bibfnamefont
  {E.}~\bibnamefont {Tafra}}, \bibinfo {author} {\bibfnamefont
  {B.}~\bibnamefont {Mihaljevi\ifmmode~\acute{c}\else \'{c}\fi{}}}, \bibinfo
  {author} {\bibfnamefont {M.}~\bibnamefont {Basleti\ifmmode~\acute{c}\else
  \'{c}\fi{}}}, \bibinfo {author} {\bibfnamefont {M.}~\bibnamefont
  {Kuve\ifmmode \check{z}\else \v{z}\fi{}di\ifmmode~\acute{c}\else
  \'{c}\fi{}}}, \bibinfo {author} {\bibfnamefont {T.}~\bibnamefont {Ivek}},
  \bibinfo {author} {\bibfnamefont {A.}~\bibnamefont
  {Hamzi\ifmmode~\acute{c}\else \'{c}\fi{}}}, \bibinfo {author} {\bibfnamefont
  {S.}~\bibnamefont {Tomi\ifmmode~\acute{c}\else \'{c}\fi{}}}, \bibinfo
  {author} {\bibfnamefont {T.}~\bibnamefont {Hiramatsu}}, \bibinfo {author}
  {\bibfnamefont {Y.}~\bibnamefont {Yoshida}}, \bibinfo {author} {\bibfnamefont
  {G.}~\bibnamefont {Saito}}, \bibinfo {author} {\bibfnamefont {J.~A.}\
  \bibnamefont {Schlueter}}, \bibinfo {author} {\bibfnamefont {M.}~\bibnamefont
  {Dressel}}, \ and\ \bibinfo {author} {\bibfnamefont {B.}~\bibnamefont
  {Korin-Hamzi\ifmmode~\acute{c}\else \'{c}\fi{}}},\ }\href {\doibase
  10.1103/PhysRevB.99.045114} {\bibfield  {journal} {\bibinfo  {journal} {Phys.
  Rev. B}\ }\textbf {\bibinfo {volume} {99}},\ \bibinfo {pages} {045114}
  (\bibinfo {year} {2019})}\BibitemShut {NoStop}%
\end{thebibliography}
%

\end{document}